# Optimization of Analytic Window Functions

Yu Cao[‡,*], Chee-Yong Chan[†], Jie Li[§,*], Kian-Lee Tan[†]

[‡]*EMC Labs, China*
yu.cao@emc.com

[†]*School of Computing, National University of Singapore, Singapore*
{chancy, tankl}@comp.nus.edu.sg

[§]*Department of Computer Science, Duke University, United States*
jieli@cs.duke.edu

## ABSTRACT

Analytic functions represent the state-of-the-art way of performing complex data analysis within a single SQL statement. In particular, an important class of analytic functions that has been frequently used in commercial systems to support OLAP and decision support applications is the class of *window functions*. A window function returns for each input tuple a value derived from applying a function over a window of neighboring tuples. However, existing window function evaluation approaches are based on a naive sorting scheme. In this paper, we study the problem of optimizing the evaluation of window functions. We propose several efficient techniques, and identify optimization opportunities that allow us to optimize the evaluation of a set of window functions. We have integrated our scheme into PostgreSQL. Our comprehensive experimental study on the TPC-DS datasets as well as synthetic datasets and queries demonstrate significant speedup over existing approaches.

## 1. INTRODUCTION

Today's mainstream commercial database systems such as DB2, Oracle and SQL Server support *analytic functions* in SQL to express complex analytical tasks. With these analytic functions, common analyses such as ranking, percentiles, moving averages and cumulative sums can be expressed concisely in a *single* SQL statement. More importantly, these functions lead to more efficient query processing - analytic queries expressed with analytic functions can potentially eliminate self-joins, correlated subqueries and/or use fewer temporary tables compared to the counterparts without such functions [19, 4]. However, to our knowledge, there were not many reported works on optimizing the processing of analytic functions.

In this paper, we focus on an important class of analytic functions, called *window functions*, that was introduced by the SQL:2003 standard and has been widely used to support OLAP and decision support applications. A window function is one of the ranking, reference, distribution and aggregate functions. However, it is evaluated over a (base or derived) table, which can be viewed as the union of logical *window partitions*. Each window partition has a single value of WPK, which is a set of attributes $\{wpk_1, wpk_2, \cdots, wpk_m\}$; each pair of window partitions are disjoint on values of WPK. With an empty WPK, the whole table forms a single window partition. In addition, tuples of each window partition are ordered by WOK, which is a sequence of attributes $(wok_1, wok_2, \cdots, wok_n)$. Within a window partition, each tuple has a *window* of neighboring tuples meeting certain criteria. The window function essentially calculates and appends a new window-function attribute to each tuple $t$, by applying a function over all tuples of the window of $t$. In other words, the evaluation of a window function wf over a table $T$ results in a new table $T'$, which contains exactly every tuple of $T$ but also has a new attribute whose values are derived by wf.

A basic *window query* block can be viewed as a normal SQL query $Q$ plus one or more window functions defined in the SELECT clause. These window functions are independent of each other. To evaluate the window query, all query clauses in $Q$ except ORDER BY and DISTINCT are first optimized and executed to derive a *windowed table*, on which these window functions are then invoked. Finally, the ORDER BY clause is imposed to sort the resultant table containing new window-function attributes to some specific order.

**Example 1** To find the rankings of each employee's salary within his department as well as the whole company, the corresponding window query defines two window functions rank_in_dept and globalrank, where the PARTITION BY key represents WPK and the ORDER BY key represents WOK:

```
SELECT empnum, dept, salary,
       rank() OVER (PARTITION BY dept ORDER BY
       salary desc nulls last) as rank_in_dept,
       rank() OVER (ORDER BY salary desc nulls
       last) as globalrank
FROM emptab;
```

In this example, the windowed table is the source table emptab. The sample output is shown below.  □

| EMPNUM | DEPT | SALARY | RANK_IN_DEPT | GLOBALRANK |
|---|---|---|---|---|
| 4 | 1 | 78000 | 1 | 3 |
| 5 | 1 | 75000 | 2 | 4 |
| 9 | 1 | 53000 | 3 | 7 |
| 7 | 2 | 51000 | 1 | 8 |
| 3 | 2 | - | 2 | 9 |
| 6 | 3 | 79000 | 1 | 2 |
| 10 | 3 | 75000 | 2 | 4 |
| 8 | 3 | 55000 | 3 | 6 |
| 2 | - | 84000 | 1 | 1 |
| 1 | - | - | 2 | 9 |

---

[*]The work was done while the authors were at National University of Singapore

In current database systems, in principle the window functions are evaluated over the windowed table as follows.

**Single window function**. Computing a single window function over the windowed table involves two logical steps. In the first step, the windowed table is reordered by a *tuple reordering operation* to physical window partitions according to the specifications of WPK and WOK. The conventional tuple reordering operation is a sort operation with a sort order given by the concatenation of some WPK's permutation and WOK. We shall refer to this reordering operation as **Full Sort (FS)**. The generated window partitions are pipelined into the second step, where the window function is sequentially invoked for each tuple within each window partition. The output table has a tuple ordering consistent with the sort order of FS.

**Multiple window functions**. A *window function chain* is formed to sequentially evaluate multiple window functions over the windowed table. The windowed table tuples are fed into the leading window function of the chain. For each of the remaining window functions, it is evaluated over the reordered output of its preceding window function. Figure 1 shows the window function chain for Example 1.

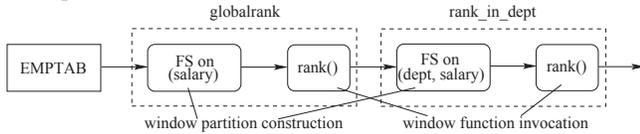

**Figure 1: A conventional window function chain for Example 1**

We note that it is possible for a certain sort order $O$ to satisfy the specifications of several consecutive window functions in the chain on their window partitions. Since a window function does not change the tuple ordering, to evaluate a sub-chain consisting of these window functions, it is sufficient to reorder only the input of the leading window function in the sub-chain to $O$ using FS. For example, consider a window function $\text{wf}_1 = (\text{WPK}_1 = \{a, b\}, \text{WOK}_1 = (c))$ followed by another window function $\text{wf}_2 = (\text{WPK}_2 = \{b\}, \text{WOK}_2 = (a))$. If the input of $\text{wf}_1$ is reordered by FS with sort order $(b, a, c)$, then $\text{wf}_2$ can be directly evaluated over the output of $\text{wf}_1$ without further tuple reordering. This *FS sharing optimization* has been adopted by some systems, e.g. Oracle [5].

In this paper, we re-examine the problem of efficient evaluation of a window function, and optimization of a chain of window functions. For a single window function evaluation, we develop two new tuple reordering mechanisms. The first method, called **Hashed Sort (HS)** is based on the key observation that window partitions delivered by a tuple reordering operation can be in an arbitrary order without affecting the result correctness and the performance of the subsequent window function invocation. Thus, under a HS operation, a table is first hashed into buckets made up of complete window partitions, and then each bucket is separately sorted to derive the physical window partitions. HS is expected to be superior over FS especially when the sorting memory is small, since it avoids a full sort.

The second method, called **Segmented Sort (SS)**, can be applied when it is possible to take advantage of the ordering of the input table. Specifically, the input of a window function should correspond to a sequence of tuple segments such that sorting each segment results in the desired window partitions for evaluating the window function. As an example, consider a window function $\text{wf}_1 = (\text{WPK}_1 = \{a\}, \text{WOK}_1 = (b))$ followed by another window function $\text{wf}_2 = (\text{WPK}_2 = \{a\}, \text{WOK}_2 = (c))$ in the chain. If the input of $\text{wf}_1$ is reordered by FS with sort order $(a, b)$ or by HS with hash key $\{a\}$ as well as sort order $(a, b)$, then the output of $\text{wf}_1$ consists of tuple segments each having a single $a$ value. As such, it is sufficient to do the tuple reordering for $\text{wf}_2$ by simply sorting each segment on $(c)$. Compared with FS and HS, SS is expected to incur a much lower cost since it does not need a full sort or table partitioning.

FS and HS have the duality analogous to that of other hash-based and sort-based query processing methods [12]. SS looks similar to the partial sort operation [7, 8, 13], which produces the required complete sort order by exploiting the partial sort order satisfied by the input. However, unlike the partial sort whose input and output both must be totally ordered, SS is more flexible as its input and output could be a *segmented relation*, which be viewed as a sequence of relation segments that is partitioned based on some attributes and sorted on some other attributes. In fact, SS can be considered as a generic and flexible extension of the partial sort operation to reorder tuples for evaluating window functions. To the best of our knowledge, we are the first to recognize and study the benefits of applying the HS and SS techniques in the context of window function evaluation.

For multiple window functions in a query, finding an optimal sequence of evaluating the window functions turns out to be NP-hard. As such, we also propose a cover set-based optimization scheme to efficiently generate a window function chain. Our scheme essentially groups the set of window functions into cover sets such that window functions within a cover set incur at most one FS/HS/SS reordering operation (for the leading window function in the cover set). We present our heuristics to partition the window functions into cover sets and to order the cover sets for processing. Our scheme naturally subsumes the *FS sharing optimization*.

Our techniques of window function evaluation can be seamlessly integrated in a typical query optimizer and work in conjunction with other complementary optimization methods (e.g., interesting orders [16] and parallel execution). We have built a prototype of our techniques within PostgreSQL [1], and conducted an extensive performance study with the TPC-DS [2] datasets as well as synthetic datasets and queries. The results showed the effectiveness of HS and SS operations over FS, and the near-optimality of our cover set-based optimization scheme over existing approaches.

The rest of this paper is organized as follows. In Section 2, we present some notations which will be utilized throughout the paper. In Section 3, we elaborate the details of the HS and SS operations. In Section 4, we describe our cover set-based optimization scheme for the evaluation of multiple window functions over a windowed table. In Section 5, we discuss how to incorporate our techniques of window function evaluation in an integrated query optimization framework. Section 6 validates the effectiveness of our proposed techniques. We discuss the related work in Section 7 and finally conclude in Section 8. Proofs of technical results are given elsewhere [9].

## 2. PRELIMINARIES

Let A be a set of attributes; let X and Y be two sequences of attributes. Besides the standard notations for sets, such as Cardinality ($|\ |$), Subset ($\subseteq, \subset$), Union ($\cup$), Interesection ($\cap$) and Complement ($-$), we also utilize the following notations:

- $\overrightarrow{A}$: a permutation of A;
- $|X|$: the number of attributes in X;
- $attr(X)$: the set of attributes in X;
- $X \circ Y$: the sequence of attributes obtained by concatenating X and Y;
- $X \wedge Y$: the longest common prefix between X and Y;
- $X(<) \leq Y$: X is a (proper) prefix of Y;
- $\varepsilon$: an empty attribute sequence.

Given a set of attributes $A$ and a relation $R$, we define $R' \subseteq R$ to a $A$-group of $R$ if $R'$ consists of all the tuples in $R$ that share the same value(s) for attribute(s) in $A$; i.e., $|\Pi_A(R')| = 1$ and $\Pi_A(R') \cap \Pi_A(R - R') = \emptyset$. Thus, $R$ is a union of $|\Pi_A(R)|$ disjoint $A$-groups.

Each window function $\text{wf}_i$ is represented by a pair $(\text{WPK}_i, \text{WOK}_i)$, where $\text{WPK}_i$ is a set of partitioning key attributes and $\text{WOK}_i$ is a sequence of ordering key attributes. For simplicity and without loss of generality, we assume that the attributes in $\text{WOK}_i$ are all ordered in ascending order.

## 3. WINDOW FUNCTION EVALUATION

In this section, we consider the evaluation of a single window function $\text{wf} = (\text{WPK}, \text{WOK})$ on a relation $R$. We first introduce a key concept termed *segmented relation* to characterize window function evaluation. Next, we present two new tuple reordering techniques, namely, Hashed Sort and Segmented Sort, to derive a segmented relation that matches a window function. We also present the cost models for these techniques and analyze their tradeoffs. Finally, we briefly describe how the execution of Hashed Sort and Segmented Sort can be parallelized for further performance improvement.

### 3.1 Segmented Relation

**Definition 1 (Segmented Relation)** *Consider a relation $R$, where $X$ is a subset of $attr(R)$, the attributes in $R$, and $Y$ is a sequence of some attribute(s) in $attr(R)$. We define $R$ to be a segmented relation w.r.t $X$ and $Y$, denoted by $R_{X,Y}$, if $R$ is ordered such that it is a sequence of $k\ (\geq 1)$ disjoint, non-empty segments, $R_1, R_2, \cdots, R_k$, that satisfy all the following properties: (1) $\bigcup_{i=1}^{k} R_i = R$; (2) the $X$ values in each pair of segments are disjoint (i.e., $\Pi_X(R_i) \cap \Pi_X(R_j) = \emptyset, \forall i, j \in [1, k], i \neq j$); and (3) each segment is sorted on $Y$.*

Note that if $X = \emptyset$, $R_{X,Y}$ is totally ordered on $Y$ and consists of exactly one segment $R$; if $X = \emptyset$ and $Y = \varepsilon$, $R_{X,Y}$ is unordered.

In general, each segment of $R_{X,Y}$ consists of one or multiple $X$-groups of $R$. For the special case where each segment $R_i$ of $R_{X,Y}$ consists of exactly one $X$-group (i.e., $|\Pi_X(R_i)| = 1$), we say that $R_{X,Y}$ is *grouped on* $X$ and denote it by $R_{X,Y}^g$. In $R_{X,Y}^g$, each segment is also ordered on every permutation $\overrightarrow{X \cup attr(Y)}$ that preserves the sequence of $attr(Y)$.

**Definition 2** *Given a segmented relation $R_{X,Y}$ and a window function $\text{wf} = (\text{WPK}, \text{WOK})$, $R_{X,Y}$ is said to match $\text{wf}$ (or $\text{wf}$ is matched by $R_{X,Y}$) if $X \subseteq \text{WPK}$ and there exists some permutation $\overrightarrow{\text{WPK}}$ of $\text{WPK}$ such that $\overrightarrow{\text{WPK}} \circ \text{WOK} \leq Y$. More generally, given a set of window functions $W$, $R_{X,Y}$ is said to match $W$ if $R_{X,Y}$ matches each $\text{wf} \in W$.*

**Example 2** Each of the segmented relations $R_{\emptyset,(a,b,c)}$, $R_{\{a\},(b,a,c)}$ and $R_{\{b\},(a,c)}^g$ matches the window function $\text{wf} = (\{a,b\},(c))$. □

A segmented relation that matches a window function has the following useful property.

**Theorem 1** *If $R$ matches a window function $\text{wf}$, then $\text{wf}$ can be evaluated on $R$ by a sequential scan of $R$ without any reordering operation.*

We explain the intuition of Theorem 1 by considering the evaluation of $\text{wf} = (\text{WPK}, \text{WOK})$ on $R_{X,Y}$. Since $X \subseteq \text{WPK}$, each segment $R_i$ of $R_{X,Y}$ consists of one or multiple $\text{WPK}$-groups. Furthermore, since each $R_i$ is ordered on $Y$ and there exists some permutation $\overrightarrow{\text{WPK}}$ such that $\overrightarrow{\text{WPK}} \circ \text{WOK} \leq Y$, each $R_i$ is necessarily also sorted on $\overrightarrow{\text{WPK}} \circ \text{WOK}$ and can be viewed as a concatenation of one or more $\text{WPK}$-groups. It follows that each $\text{WPK}$-group in $R_i$ is ordered on $\text{WOK}$. Thus, $\text{wf}$ can be evaluated by a sequential scan of the sequence of ordered $\text{WPK}$-groups in $R_{X,Y}$.

Based on Theorem 1, to evaluate a window function $\text{wf} = (\text{WPK}, \text{WOK})$ on $R$, it suffices to reorder $R$ (if $R$ does not match $\text{wf}$) to obtain a $R_{X,Y}$ that matches $\text{wf}$ and then sequentially scan $R_{X,Y}$. The most straightforward approach to achieve this reordering is the Full Sort ($\text{FS}$) technique which sorts $R$ on $\overrightarrow{\text{WPK}} \circ \text{WOK}$ for some permutation $\overrightarrow{\text{WPK}}$ of $\text{WPK}$. The sorted result $R'$, which is essentially $R_{\emptyset, \overrightarrow{\text{WPK}} \circ \text{WOK}}$, trivially matches $\text{wf}$. However, note that the total ordering of $R'$ on $\overrightarrow{\text{WPK}} \circ \text{WOK}$ is actually unnecessary for the purpose of computing $\text{wf}$, which only requires the input tuples to be partially sorted (i.e., a sequence of tuple partitions grouped on $\text{WPK}$ and then each sorted on $\text{WOK}$).

Before we present more efficient reordering techniques in the next two sections, we first introduce the notion of reorderability.

**Definition 3 (Reorderable)** *Given a window function $\text{wf}$, a relation $R$, and a reordering technique $O$, we say that $(R, \text{wf})$ is $O$-reorderable if $R$ could be reordered by $O$ such that the reordered relation matches $\text{wf}$. More generally, given a set of window functions $W$, we say that $(R, W)$ is $O$-reorderable if $(R, \text{wf})$ is $O$-reorderable for each $\text{wf} \in W$.*

### 3.2 Hashed Sort Technique

Hashed Sort ($\text{HS}$) reorders $R$ wrt $\text{wf}$ in two steps: the first step partitions $R$ into a collection of buckets by hashing $R$ on some hash key, $\text{WHK} \subseteq \text{WPK}$, and the second step sorts each bucket $R_i$ on a sort key $\overrightarrow{\text{WPK}} \circ \text{WOK}$ for some permutation $\overrightarrow{\text{WPK}}$ of $\text{WPK}$. Thus, $\text{HS}$ essentially reorders $R$ to obtain $R_{\text{WHK}, \overrightarrow{\text{WPK}} \circ \text{WOK}}$ which matches $\text{wf}$. In order that $\text{HS}$ does not degenerate to $\text{FS}$, we require that $\text{WHK} \neq \emptyset$; thus, $(R, \text{wf})$ is $\text{HS}$-reorderable if $\text{WPK} \neq \emptyset$.

**Example 3** $R$ can be reordered by $\text{HS}$ to match $\text{wf} = (\{a,b\},(c))$ if $\text{WHK}$ is $\{a\}, \{b\}$ or $\{a,b\}$, and $\overrightarrow{\text{WPK}}$ is $(a,b)$ or $(b,a)$. □

The details of $\text{HS}$ are as follows. The first step sequentially scans $R$ to build the buckets by hashing on $\text{WHK}$. $\text{HS}$ tries to maintain as many buckets resident in the allocated main-memory as possible. Whenever the memory is full, $\text{HS}$ picks a bucket $R_i$ to be flushed to disk, and any subsequent tuple for $R_i$ will be flushed to disk. At the end of the partitioning step, some of the buckets are resident in main-memory while the remaining ones are resident on disk. The second step will first sort the memory-resident buckets before the disk-resident ones.

$\text{HS}$ can be further optimized as follows. If statistics on the $\text{WHK}$ values are available (e.g., histograms on a base relation $R$), it is possible to estimate the most frequent $\text{WHK}$ values (MFVs), each of which corresponds to a set of tuples whose total size exceeds the size of sorting memory. Tuples with such values belong a special bucket $R_x$ that will be immediately pipelined for sorting (i.e., without being cached in main memory or flushed to disk in contrast to other tuples). Therefore, $R_x$ is sorted before any other bucket. Such an optimization could save up to one pass of I/O for $R_x$. Moreover, it is likely to result in a larger set of in-memory hashed buckets which can be sorted internally.

### 3.3 Segmented Sort Technique

Segmented Sort ($\text{SS}$) is designed to reorder a relation $R_{X,Y}$ to match a window function $\text{wf} = (\text{WPK}, \text{WOK})$. As shown below, $\text{SS}$ performs the reordering by separately sorting each segment/group of $R_{X,Y}$, whose size is generally much smaller than the entire relation. Thus, SS is usually much more efficient than FS and HS.

$(R_{X,Y}, \texttt{wf})$ is SS-reorderable if one of the following conditions hold: either (1) $X \neq \emptyset$ and $X \subseteq \texttt{WPK}$, or (2) $X = \emptyset$ and there exists some permutation $\overrightarrow{\texttt{WPK}}$ of WPK such that $(\overrightarrow{\texttt{WPK}} \circ \texttt{WOK}) \wedge Y$ is non-empty. Specifically, SS reorders $R_{X,Y}$ to $R_{X,\overrightarrow{\texttt{WPK}} \circ \texttt{WOK}}$ for some permutation $\overrightarrow{\texttt{WPK}}$ of WPK. Note that if $X = \emptyset$, we need to choose a permutation $\overrightarrow{\texttt{WPK}}$ such that $(\overrightarrow{\texttt{WPK}} \circ \texttt{WOK}) \wedge Y$ is non-empty; this permutation must exist by the applicability requirement of SS. As we shall explain later, the constraint imposed on $\overrightarrow{\texttt{WPK}}$ when $X = \emptyset$ is to ensure that SS does not degenerate to FS, which requires sorting the entire $R_{X,Y}$. By Definition 2, it follows that the reordered relation $R_{X,\overrightarrow{\texttt{WPK}} \circ \texttt{WOK}}$ matches wf since $X \subseteq \texttt{WPK}$.

We now explain how $R_{X,\overrightarrow{\texttt{WPK}} \circ \texttt{WOK}}$ is derived from $R_{X,Y}$. Let $\alpha = (\overrightarrow{\texttt{WPK}} \circ \texttt{WOK}) \wedge Y$. Thus, $\overrightarrow{\texttt{WPK}} \circ \texttt{WOK} = \alpha \circ \beta$, where $\beta$ is some sequence of attribute(s)[1]. There are two cases to consider depending on whether $\alpha$ is empty. Consider the general case where $\alpha$ is non-empty. Since $\alpha \leq Y$, therefore each segment $R_i$ of $R_{X,Y}$, which is ordered on $Y$, is necessarily also ordered on $\alpha$. Thus, each segment $R_i$ is actually a sequence of $\alpha$-groups. By separately sorting each of these $\alpha$-groups on $\beta$, each segment $R_i$ becomes sorted on $\alpha \circ \beta = \overrightarrow{\texttt{WPK}} \circ \texttt{WOK}$. Thus, we have reordered $R_{X,Y}$ to $R_{X,\overrightarrow{\texttt{WPK}} \circ \texttt{WOK}}$.

**Example 4** Consider using SS to reorder $R$ to $R'$ wrt $\texttt{wf} = (\{a,b\}, (c))$. If $R = R_{\emptyset,(a,d)}$, then $\alpha = (a)$ and $R' = R_{\emptyset,(a,b,c)}$. If $R = R_{\{a\},(a,b,d)}$, then $\alpha = (a,b)$ and $R' = R_{\{a\},(a,b,c)}$. If $R = R^g_{\{b\},(a,d)}$, then $\alpha = (a,b)$ and $R' = R^g_{\{b\},(a,c)}$. □

Consider the second case where $\alpha$ is empty (i.e., $\beta = \overrightarrow{\texttt{WPK}} \circ \texttt{WOK}$). By separately sorting each segment $R_i$ of $R_{X,Y}$ on $\beta$, we have reordered $R_{X,Y}$ to $R_{X,\overrightarrow{\texttt{WPK}} \circ \texttt{WOK}}$. Note that by the applicability requirement of SS, $\alpha$ is empty necessarily implies that $X \neq \emptyset$ as otherwise, we would have selected a permutatation of WPK that guarantees that $\alpha$ is non-empty. As we alluded earlier, the reason to avoid having both $X$ and $\alpha$ being empty is to ensure that SS does not degenerate to FS: if $X$ were empty (i.e., $R_{X,Y}$ consists of a single segment), then SS would essentially be performing a complete sort of the entire $R_{X,Y}$ on $\beta$.

**Example 5** Consider two examples of using SS to reorder $R$ to $R'$ wrt $\texttt{wf} = (\{a,b\}, (c))$. If $R = R_{\{a\},(d)}$ then $R' = R_{\{a\},(a,b,c)}$. If $R = R_{\{b\},(c)}$ then $R' = R_{\{b\},(a,b,c)}$. □

Observe that there is generally more than one way for SS to reorder $R_{X,Y}$ to match wf depending on the choice of $\overrightarrow{\texttt{WPK}}$. For the general case where $\alpha$ is non-empty, it makes sense to choose the permutation of WPK that maximizes the total number of distinct values of $\alpha$[2] so that the size of each $\alpha$-group within each segment is minimized resulting in more efficient sorting and hence reordering.

We remark that the partial sort operation [7, 13] is essentially an instance of SS, where the input relation is $R_{\emptyset,Y}$, the window function to match is $\texttt{wf} = (\emptyset, \texttt{WOK})$ and $Y < \texttt{WOK}$. Clearly, SS has much broader applicability than the partial sort.

We conclude the discussion on SS by presenting a useful property (cf. [9] for the proof due to space limitation) for reasoning about SS-reorderability.

**Theorem 2** *Let $\texttt{wf}_1$ and $\texttt{wf}_2$ be two distinct window functions, and let $R$ be a relation.*

1. *If $R$ matches $\texttt{wf}_1$ and $R'$ is the output produced by evaluating $\texttt{wf}_1$ on $R$, then $(R, \texttt{wf}_2)$ is SS-reorderable iff $(R', \texttt{wf}_2)$ is SS-reorderable.*
2. *If $(R, \texttt{wf}_1)$ is SS-reorderable and $R'$ is produced by reordering $R$ with SS to match $\texttt{wf}_1$, then $(R, \texttt{wf}_2)$ is SS-reorderable iff $(R', \texttt{wf}_2)$ is SS-reorderable.*

Theorem 2 essentially states that the SS-reorderability property of a relation $R$ (wrt some window function $\texttt{wf}_2$) is preserved by two types of transformation of $R$ to $R'$: (1) evaluating some window function $\texttt{wf}_1$ on $R$ to obtain $R'$, and (2) reordering $R$ (wrt some window function $\texttt{wf}_1$) using SS to obtain $R'$. By preservation, we mean that $(R, \texttt{wf}_2)$ is SS-reorderable iff $(R', \texttt{wf}_2)$ is SS-reorderable. This property will be used in our optimization framework in Section 4.

### 3.4 Cost Models and Analysis

In this section, we present cost models for reordering a relation $R$ (of the form $R_{X,Y}$) to match $\texttt{wf} = (\texttt{WPK}, \texttt{WOK})$ using the reordering operators FS, HS and SS. Since a reordering operator may pipeline its output (to another operator) while the reordering is still in progress, our cost models include the cost of outputting the reordered relation but exclude the cost of reading the input relation.

We use $Cost(R, O)$ to denote the cost of reordering $R$ using operator $O$, $M$ to denote the allocated main memory (in number of blocks) for the operation, and $B(R_i)$ to denote the size of a relation/segment/group $R_i$ (in number of blocks). Let $k$ denote the number of segments in $R_{X,Y}$.

FS is based on the standard, external merge-sort algorithm consisting of two phases: an initial run formation phase that creates sorted subsets, called runs, and a merge phase that merges runs into larger runs iteratively, until a single run is created. Assuming that replacement selection is used to create the initial sorted runs, the size of each initial run is $2M$ blocks. The sorted runs are merged using the well-known $F$-way merge pattern, where $F$ is the merge order (i.e., number of runs that can be simultaneously merged using $M$). Therefore,

$$Cost(R, \texttt{FS}) = 2 \times B(R) \times (\lceil log_F(\frac{B(R)}{2M}) \rceil + 1) \quad (1)$$

For HS, we assume that the values of WHK follow a uniform distribution. If the number of distinct values of WHK in $R$, denoted by $D(\texttt{WHK})$, is large enough, we expect that each generated hashed bucket will be small enough to fit into main memory and thus, it can be internally sorted; otherwise, if $D(\texttt{WHK})$ is very small, the hashed buckets may require external sortings. As such, we estimate the total number of generated hashed buckets as $N = D(\texttt{WHK})$. Therefore, $B(R_i)$ is estimated as $B(R)/N$ for each hashed bucket $R_i$. The number of hashed buckets that are never flushed to disk is $N' = \lfloor M \times N/B(R) \rfloor$. Therefore,

$$Cost(R, \texttt{HS}) = 2 \times B(R) \times (1 - \frac{N'}{N}) + \sum_{i=1}^{N} Cost(R_i) \quad (2)$$

where $Cost(R_i)$ denotes the cost of (internally or externally) sorting the $i^{th}$ hashed bucket $R_i$. Since the sortings in HS incur possibly less I/O cost (due to possibly fewer run merge passes) than FS, $\sum_{i=1}^{N} Cost(R_i)$ is expected to be lower than $Cost(R, \texttt{FS})$. As such, $Cost(R, \texttt{HS})$ is lower than $Cost(R, \texttt{FS})$ when the value of $Cost(R, \texttt{FS}) - \sum_{i=1}^{N} Cost(R_i)$ is large enough, i.e., when $M$ is small. Thus, we expect that HS is generally comparable to FS, but HS will outperform FS when $M$ is small.

For SS, recall from Section 3.3 that SS reorders by independently sorting either segments of $R_{X,Y}$ if $\alpha$ is empty; or $\alpha$-groups within each segment of $R$, otherwise. For convenience, we refer to each segment/group being sorted as a *unit*. To model the sorting cost, we need to estimate the number and size of the units. Let $u$ denote the number of units in each segment of $R$. We assume that the attributes of $R$ follow the uniform distribution and are uncorrelated with each other. As such, for each segment $R_i$, $B(R_i) = B(R)/k$.

---
[1] Note that since $R_{X,Y}$ does not match wf, $\alpha \neq \overrightarrow{\texttt{WPK}} \circ \texttt{WOK}$. Therefore, $\beta$ contains at least one attribute.
[2] Since $Y$ is fixed for a given $R_{X,Y}$, this translates to maximizing the number of attributes in $\alpha$.

There are two cases to consider depending on whether $\alpha$ is empty. For the case where $\alpha$ is empty, each unit is a segment and $u = 1$.

We now consider the case where $\alpha$ is non-empty. Note that each segment contains a proper subset of the distinct $X$ values in $R$. If a segment is large enough, we assume that it contains all of (resp. $1/k$ of) the distinct $\alpha$ values in $R$ when $attr(\alpha) \cap X$ is empty (resp. non-empty); otherwise, we assume that each tuple in the segment has a distinct $\alpha$ value. Therefore,

$$u = \begin{cases} min(T(R)/k, D(\alpha)) & \text{if } attr(\alpha) \cap X = \emptyset \\ min(T(R)/k, D(\alpha)/k) & \text{otherwise.} \end{cases}$$

where $T(R)$ denotes the number of tuples in $R$ and $D(\alpha)$ denotes the number of distinct values of $\alpha$ in $R$. Thus, $R$ contains a total of $k * u$ units, each of which has a size of $B(R)/(k * u)$ blocks. Therefore,

$$Cost(R, \text{SS}) = \sum_{i=1}^{k*u} Cost(U_i) \quad (3)$$

where $Cost(U_i)$ denotes the cost of (internally or externally) sorting a unit $U_i$.

SS can be very efficient without incurring much or any I/O overhead, especially when the units to be sorted are small. Comparing Eqs. 2 and 3, SS is at least no more expensive than HS. Moreover, compared with FS, SS has significantly fewer number of tuple comparisons without incurring extra I/O cost: the complexity of independently sorting $k$ segments each of $n/k$ tuples is $O(k * n/k log(n/k)) = O(n log(n/k))$ compared to a complexity of $O(n log(n))$ for a single sort of all $n$ tuples. Thus, the cost of SS is expected to be generally lower than FS and HS.

### 3.5 Parallel Execution

The evaluation of a window function $\text{wf} = (\text{WPK}, \text{WOK})$ on a relation $R$ can be easily parallelized by partitioning the tuples of $R$ by either hash or range partitioning on the WPK attributes. The window function wf can then be evaluated in parallel on each data partition. It is easy to see that if $(R, \text{wf})$ is SS-reorderable (resp. HS-reorderable), then each data partition can also be processed by reordering its tuples with an appropriate SS (resp. HS) operation.

## 4. OPTIMIZATION OF MULTIPLE WINDOW FUNCTION EVALUATIONS

In this section, we consider the general problem of evaluating a set of window functions in a query. Specifically, the problem is to optimize the evaluation of a set of window functions $W = \{\text{wf}_1, \cdots, \text{wf}_n\}$ on a relation $R$, where each $\text{wf}_i = (\text{WPK}_i, \text{WOK}_i)$ and $R$ is of the form $R_{X,Y}$ for some set of attributes $X$ and some sequence of attributes $Y$. Recall that if $R$ is an unordered relation, then $X$ is an empty set and $Y$ is an empty sequence.

### 4.1 Evaluation Model

The window functions in $W$ are evaluated sequentially based on some ordering of the window functions. Let $(\text{wf}_1, \cdots, \text{wf}_n)$ denote the chosen evaluation order, and let $I_j$ and $O_j$ denote, respectively, the input and output relations of the evaluation of each $\text{wf}_j \in W$. Each window function $\text{wf}_i$ is evaluated by the following two steps. First, if $I_j$ does not match $\text{wf}_j$, then reorder $I_j$ to $I'_j$ using an applicable reordering technique (i.e., FS/HS/SS) such that $I'_j$ matches $\text{wf}_j$. For convenience, let $I'_j$ denote $I_j$ if there is no reordering. Second, sequentially scan $I'_j$ to compute $\text{wf}_j$. Note that $I_j$ is the original relation $R_{X,Y}$ if $j = 1$; otherwise, $I_j$ is $O_{j-1}$.

The above sequential evaluation model is implemented in several database systems including DB2, Oracle, SQL Server[3] and PostgreSQL.

To optimize the evaluation of $W$, we need to choose an evaluation order of the window functions and choose a reordering technique for each window function that is not matched by its input relation. The following result establishes that this optimization problem is NP-hard.

**Theorem 3** *The problem of finding the lowest-cost evaluation plan for an input set of window functions is NP-hard.*

The proof is established by reducing the Travelling Salesman Problem [14] to a special case of the problem, where a compulsive FS will be used to reorder the input of every window function (cf. [9] for the complete proof).

### 4.2 Overview of Our Approach

Given the NP-hardness of optimizing a set of window function evaluations, in this paper, we present an efficient heuristic to solve the problem. Our approach optimizes the evaluation of $W$ by minimizing two key aspects: (1) the number of reorder operations, and (2) the usage of FS and HS (which are generally less efficient than SS) for reordering.

Note that as each window function evaluation computes an additional column to store the derived values for some analytic function, the size of the input relation for each window function evaluation actually becomes larger as the evaluation progresses. However, for tractability reasons, our optimization framework makes a simplifying assumption that the size of the input and output relations for each window function evaluation are the same. We refer to this as *relation size assumption*. As the number of window functions is not too many, the additional columns introduced by the window function evaluations are relatively small compared to the tuple size. In Section 4.6, we discuss how our approach can be further optimized to mitigate this assumption. As we shall see in the experimental results, our optimization framework is effective even with this simplifying assumption.

The following two examples illustrate the intuitions for our optimization framework.

**Example 6** Consider the evaluation of $W = \{\text{wf}_1 = (\{a\}, (b)), \text{wf}_2 = (\{a\}, \varepsilon)\}$ on an input relation $R_{\emptyset,\varepsilon}$, which matches none of $\text{wf}_1$ and $\text{wf}_2$. If we first reorder $R$ into $R_{\emptyset,(a,b)}$ for evaluating $\text{wf}_1$, then the output of $\text{wf}_1$ directly matches $\text{wf}_2$. In contrast, if we first reorder $R$ into $R_{\emptyset,(a)}$ for evaluating $\text{wf}_2$, then we need an extra SS operation to reorder the output of $\text{wf}_2$ for evaluating $\text{wf}_1$. □

**Example 7** Consider the evaluation of $W = \{\text{wf}_1 = (\{a, b\}, \varepsilon), \text{wf}_2 = (\{a\}, (c))\}$ on an input relation $R_{\emptyset,\varepsilon}$, which matches none of $\text{wf}_1$ and $\text{wf}_2$. Suppose we first reorder $R$ with a FS operation for evaluating $\text{wf}_1$. If $R$ is reordered to $R_{\emptyset,(a,b)}$, then we just need a SS operation to reorder the output of $\text{wf}_1$ for evaluating $\text{wf}_2$. On the other hand, if $R$ is reordered to $R_{\emptyset,(b,a)}$, then we need a more expensive FS/HS operation to reorder the output of $\text{wf}_1$ for $\text{wf}_2$. □

As illustrated by Example 6, a useful strategy to reduce the number of reorder operations is to identify a subset $W_i$ of window functions that can be matched by a common segmented relation $R_i$. The idea is that instead of incurring possibly one reorder operation to evaluate each window function in $W_i$, we can just perform a single reordering of the input relation to derive $R_i$ which can then be used to evaluate $W_i$. In the following, we formalize the required properties for the above evaluation idea.

---

[3]For the commerical DBMS, our conclusions are drawn from viewing the physical plans of our test queries. We note that Oracle also supports parallel evaluation of a single window function [4, 5].

**Theorem 4** *Let $R$ be a relation on which a set of window functions $W$ are evaluated. If $R$ matches $W$, then for any evaluation order $(\mathtt{wf}_1, \cdots, \mathtt{wf}_n)$ of the window functions in $W$, the output relation $O_i$ (produced by the evaluation of $\mathtt{wf}_i$ on $I_i$) matches $W$ for each $\mathtt{wf}_i \in W$.*

**Corollary 1** *If $R$ matches a set of window functions $W$, then for any evaluation order of $W$, $W$ can be evaluated on $R$ without any reordering operation.*

Thus, by Corollary 1 (which is a generalization of Theorem 1), if a set of window functions $W$ is not matched by a relation $R$ and it is possible to reorder $R$ to $R'$ so that $R'$ matches $W$, then we can evaluate $W$ on $R$ by a single reordering operation. Specifically, for any evaluation order $(\mathtt{wf}_1, \cdots, \mathtt{wf}_n)$ of $W$, the evaluation of $\mathtt{wf}_1$ requires a reordering of $R$ to $R'$; subsequently, the evaluation of each $\mathtt{wf}_i$, $i > 1$, does not require any reordering.

We next characterize an important property for a relation to match a set of window functions.

**Definition 4** *A set of window functions $W$ is defined to be a cover set if there exists a permutation $\overrightarrow{\mathtt{WPK}_i}$ of $\mathtt{WPK}_i$ for each window function $\mathtt{wf}_i \in W$ and a window function $\mathtt{wf}_c \in W$ such that $\overrightarrow{\mathtt{WPK}_i} \circ \mathtt{WOK}_i \le \overrightarrow{\mathtt{WPK}_c} \circ \mathtt{WOK}_c$ for each $\mathtt{wf}_i \in W - \{\mathtt{wf}_c\}$. The window function $\mathtt{wf}_c$ is defined to be a covering window function of $W$, and $\overrightarrow{\mathtt{WPK}_c} \circ \mathtt{WOK}_c$ is defined to be a covering permutation of $\mathtt{wf}_c$.*

**Example 8** Consider $W = \{\mathtt{wf}_1, \mathtt{wf}_2, \mathtt{wf}_3\}$, where $\mathtt{wf}_1 = (\{a, b, c\}, (d))$, $\mathtt{wf}_2 = (\{a, b\}, (c, d))$, and $\mathtt{wf}_3 = (\{a, b\}, (c))$. $W$ is a cover set with two covering window functions $\mathtt{wf}_1$ and $\mathtt{wf}_2$. □

**Theorem 5** *If a relation $R$ matches a set of window functions $W$, then $W$ is a cover set.*

Theorem 5 (cf. [9] for the proof) suggests the following idea to optimize the evaluation of $W$. Let $W$ be partitioned into a collection of cover sets, $W = C_0 \cup \cdots \cup C_k$, such that each $C_i$ is evaluated before $C_{i+1}$.

For each cover set $C_i$, if the input relation to $C_i$ either matches the first window function in $C_i$ or can be reordered to some relation that matches the first window in $C_i$, then it follows that each $C_i$ can be evaluated by at most one reordering operation, and therefore, $W$ can be evaluated using at most $(k + 1)$ reorderings. Thus, by minimizing the number of cover sets in the partitioning of $W$, the number of reorder operations to evaluate $W$ can be minimized.

Our evaluation approach builds on the above idea to evaluate $W$ as a sequence of *cover set evaluations*, where each cover set evaluation is a sequence of window function evaluations. We elaborate on this cover set-based evaluation strategy in the next section.

### 4.3 Cover Set-based Evaluation

Before we present our approach, we shall introduce some notations. For each cover set $C_i$, let $\mathcal{I}_i$ and $\mathcal{O}_i$ denote, respectively, the input and output relations of the evaluation of $C_i$; i.e. $\mathcal{I}_i$ is the input relation to the first window function in $C_i$, and $\mathcal{O}_i$ is the output relation produced by the last window function in $C_i$. Let $\mathtt{wf}_i^*$ denote the first window function that is evaluated in cover set $C_i$.

To minimize both the number of reorder operations as well as the number of reorderings performed using FS/HS, our approach partitions $W$ into three disjoint subsets, $W = C_0 \cup C_1 \cup C_2$, such that $C_0$ is evaluated first, followed by $C_1$, and finally $C_2$. Note that each $C_i$ could possibly be empty.

$C_0$ is the set of window functions in $W$ that are matched by the input relation $R_{X,Y}$. Thus, $C_0$ is necessarily a cover set (by Theorem 5), and by Corollary 1, each window function in $C_0$ can be evaluated without any reordering.

Since $R_{X,Y}$ does not match any $\mathtt{wf}_i \in W - C_0$, the remaining set of window functions (i.e., $C_1 \cup C_2$) requires at least one reordering to be evaluated. To minimize the usage of FS/HS reorderings, $C_1$ is defined to contain all the window functions in $W - C_0$ such that $(R_{X,Y}, C_1)$ is SS-reorderable. By Theorem 2, $(\mathcal{O}_0, C_1)$ is necessarily also SS-reorderable, where $\mathcal{O}_0$ is the output relation produced by the evaluation of $C_0$. Thus, $C_1$ can be evaluated using only SS reorderings (i.e., FS/HS reorderings can be avoided). To minimize the number of SS reorderings to evaluate $C_1$, our approach further partitions $C_1$ into a minimum number of disjoint cover sets, $C_1 = C_{1,1} \cup \cdots \cup C_{1,m_1}$. The details are explained in Section 4.4.

The evaluation of $C_2$, however, is more intricate as it requires at least one FS/HS reordering and zero or more SS reorderings. To minimize the number of reorderings to evaluate $C_2$, $C_2$ is also evaluated in terms of a collection of cover sets, $C_2 = (C_{2,1} \cup \cdots \cup C_{2,m_2}) \cup \cdots \cup (C_{k,1} \cup \cdots \cup C_{k,m_k})$. The details are explained in Section 4.5.

The overall organization of our cover set-based evaluation is depicted in Figure 2. Each circle in Figure 2 represents a single window function evaluation, where the color indicates the technique used for reordering (if any): white means that there is no reordering, and gray (black, resp.) means that the input relation is reordered using SS (FS or HS, resp.). The window functions within each box represent a cover set, and the chain of window function evaluations are connected by the directed edges. Except for $C_0$, which is evaluated without any reordering, each of the cover sets $C_{i,j}$ requires exactly one reordering for its evaluation, and the reordering is performed as part of the evaluation of the first window function.

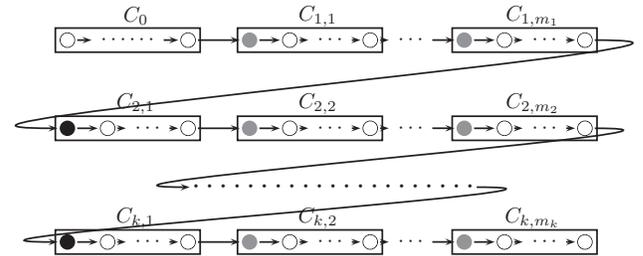

**Figure 2: Cover set-based evaluation approach**

In the following, we discuss the optimization of the evaluations of $C_1$ and $C_2$.

### 4.4 Evaluation of $C_1$

As explained in the previous section, $(\mathcal{O}_0, C_1)$ is SS-reorderable, and the number of required SS reorderings to evaluate $C_1$ is minimized by partitioning $C_1$ into a minumum number of cover sets. However, the problem of finding such an optimal partitioning of $C_1$ is NP-hard as the following result shows.

**Theorem 6** *The problem of partitioning a set of window functions $W$ into a minimum number of cover sets is NP-hard.*

The proof is established by a reduction from the Minimum Vertex Coloring Problem [14] (cf. [9] for the complete proof).

As such, the partitioning of $C_1$ can be solved using an efficient heuristic (e.g., Brelaz's heuristic algorithm [6] for the minimum vertex coloring problem).

1249

Assume that $C_1$ has been partitioned into $m_1$ cover sets, $C_1 = C_{1,1} \cup \cdots \cup C_{1,m_1}$, and the evaluation order is $C_{1,1}, \cdots, C_{1,m_1}$. Thus, $\mathcal{I}_{1,1} = \mathcal{O}_0$. Since $(\mathcal{O}_0, C_1)$ is SS-reorderable, it follows from Theorem 2 that for each $C_{1,j}$, $(\mathcal{I}_{1,j}, C_{1,j})$ is SS-reorderable.

Each $C_{1,j}$ can be evaluated on $\mathcal{I}_{1,j}$ using a single SS-reordering based on the following result (cf. [9] for the proof).

**Theorem 7** *Consider the evaluation of a set of window functions $W$ on a relation $R_{X,Y}$, where $W$ is a cover set and $(R, W)$ is SS-reorderable. Let $R_{X,Y'}$ be the output relation produced by a SS reordering of $R$ wrt a covering function $\mathtt{wf}_c$ of $W$ such that $Y'$ is a covering permutation of $\mathtt{wf}_c$. Then $R_{X,Y'}$ matches $W$.*

Let $\mathcal{I}_{1,j}$ be of the form $R_{X,Y}$. To apply Theorem 7 to evaluate $C_{1,j}$ on $\mathcal{I}_{1,j}$, we choose a covering function of $C_{1,j}$ to be $\mathtt{wf}_{1,j}^*$, the first window function to be evaluated in $C_{1,j}$. Since $\mathcal{I}_{1,j}$ does not match $\mathtt{wf}_{1,j}^*$ but $(\mathcal{I}_{1,j}, \mathtt{wf}_{1,j}^*)$ is SS-reorderable, we reorder $\mathcal{I}_{1,j}$ to $\mathcal{I}'_{1,j}$ using SS wrt $\mathtt{wf}_{1,j}^*$ such that $\mathcal{I}'_{1,j}$ is of the form $R_{X,Y'}$, where $Y'$ is a covering permutation of $\mathtt{wf}_{1,j}^*$. By Theorem 7, $\mathcal{I}'_{1,j}$ matches $C_{1,j}$, and therefore each $\mathtt{wf}_i \in C_{1,j}$ can be evaluated without any further reordering.

## 4.5 Evaluation of $C_2$

By definition of $C_2$, for each $\mathtt{wf}_i \in C_2$, $\mathtt{wf}_i$ is not matched by $R_{X,Y}$ and $(R_{X,Y}, \mathtt{wf}_i)$ is not SS-reorderable. Furthermore, it follows from Theorem 2 that $(\mathcal{O}_{1,m_1}, \mathtt{wf}_i)$ is also not SS-reorderable, where $\mathcal{O}_{1,m_1}$ is the output relation produced by the evaluation of $C_1$. Therefore, the evaluation of $C_2$ requires at least one reordering using FS/HS.

To minimize the number of FS/HS reorderings to evaluate $C_2$, we partition $C_2$ into a minimum number of partitions, $C_2 = P_2 \cup \cdots \cup P_k$[4], such that each $P_i$ can be evaluated with exactly one FS/HS reordering and zero or more SS reorderings. To minimize the number of SS reorderings required for evaluating each $P_i$, we further partition each $P_i$ into a minimum collection of cover sets, $P_i = C_{i,1} \cup \cdots \cup C_{i,m_i}$.

The collection of cover sets in $C_2$ are evaluated in the following order: each $P_i$ is evaluated before $P_{i+1}$, and within each $P_i$, each $C_{i,j}$ is evaluated before $C_{i,j+1}$. The entire order of cover set evaluations is shown in Figure 2.

Note that within each $P_i$, it is necessary for the first cover set $C_{i,1}$ to be reordered using FS/HS. This is a consequence of the fact that $(R_{X,Y}, \mathtt{wf})$ is not SS-reorderable for each $\mathtt{wf} \in C_2$. Thus, within each $P_i$, $C_{i,1}$ is reordered using FS/HS, while each of the remaining cover sets $C_{i,j}, j > 1$, in $P_i$ is reordered using SS.

For the above evaluation strategy for $C_2$ to be feasible, it is necessary that $(\mathcal{I}_{i,j}, C_{i,j})$ is SS-reorderable for each $P_i$ in $C_2$ and for each $j \in [2, m_i]$, so that each of the cover sets in $P_i$ (except for the first) can be reordered using SS. The following result states the required property for this strategy to work.

**Definition 5 (Prefixable)** *A set of window functions $W = \{\mathtt{wf}_1, \cdots, \mathtt{wf}_n\}$ is defined to be prefixable if for each $\mathtt{wf}_i \in W$, there exists a permutation $\overrightarrow{\mathtt{WPK}_i}$ of $\mathtt{WPK}_i$ such that $\bigwedge_{i=1}^{n}(\overrightarrow{\mathtt{WPK}_i} \circ \mathtt{WOK}_i)$ is non-empty.*

**Theorem 8** *Let a set of window functions $W$ be evaluated on a relation $R$, where for each $\mathtt{wf}_i \in W$, $R$ does not match $\mathtt{wf}_i$ and $(R, \mathtt{wf}_i)$ is not SS-reorderable. $W$ can be evaluated with one FS/HS reordering and zero or more SS reorderings iff $W$ is prefixable.*

---
[4]For notational convenience, we label the partitions of $C_2$ to start from $P_2$ instead of $P_1$. As each $P_i$ is further partitioned into cover sets $C_{i,j}$, this ensures that the cover sets of $C_2$ are distinctly labeled from those of $C_1$.

Based on Theorem 8 (cf. [9] for the proof), our evaluation strategy for $C_2$ requires that each $P_i$ be prefixable. However, the problem of finding such an optimal partitioning of $C_2$ is NP-hard as the following result shows.

**Theorem 9** *The problem of partitioning a set of window functions $W$ into a minimum number of prefixable, disjoint subsets is NP-hard.*

The proof is established by reducing the Minimum Set Cover problem [14] to a special case of the problem (cf. [9] for the complete proof). The partitioning problem can be solved using a greedy heuristic that tries to minimize the number of prefixable subsets by maximizing the number of window functions in each prefixable subset under construction. The details of the heuristic are given elsewhere [9] and it has $O(|W|^2)$ time-complexity. The effectiveness of the heuristic has been validated by our experimental results in Section 6.2, where it succeeded in finding the optimal partitioning of $C_2$ for all tested window queries.

Assume that $C_2$ has been partitioned into $k$ prefixable subsets, $C_2 = P_2 \cup \cdots \cup P_k$, and each $P_i$ has been partitioned into $m_i$ cover sets, $P_i = C_{i,1} \cup \cdots \cup C_{i,m_i}$ as discussed.

Each $P_i$ is processed using two main steps. In the first step, we reorder $\mathcal{I}_{i,1}$ (wrt $\mathtt{wf}_{i,1}^*$ in $C_{i,1}$) to $\mathcal{I}'_{i,1}$ such that the following two properties are satisfied: (1) $C_{i,1}$ can be evaluated with exactly one reordering (using FS/HS); and (2) each of the remaining $C_{i,j}$ in $P_i$ can be evaluated with exactly one reordering (using SS).

For this reordering operation, if both FS and HS are applicable, the choice of which technique to apply is determined in a cost-based manner. We discuss these two cases of reordering in the following two subsections.

The second step evaluates each $C_{i,j}$ as follows. By Property 1, we use $\mathcal{I}'_{i,1}$ to evaluate $C_{i,1}$ without any further reorderings. By Property 2, we use $\mathcal{I}'_{i,1}$ to evaluate $P_i - \{C_{i,1}\}$ following the same procedure as evaluating $C_1$ using $\mathcal{O}_0$.

In the ensuing discussion, we will present the details of the first step of reordering $\mathcal{I}_{i,1}$ using FS/HS.

### 4.5.1 Reordering with FS

We first discuss how to reorder $\mathcal{I}_{i,1}$ (wrt $\mathtt{wf}_{i,1}^*$) to $\mathcal{I}'_{i,1}$ using FS. The main task is to determine the sort key for FS such that it satisfies two properties: (1) $\mathcal{I}'_{i,1}$ matches $C_{i,1}$, and (2) $(\mathcal{I}'_{i,1}, \mathtt{wf}_{i,j}^*)$ is SS-reorderable for each $j \in [2, m_i]$. Property 1 ensures that $C_{i,1}$ can be evaluated using exactly one reordering with FS, and Property 2 ensures that each of the remaining $C_{i,j}$ in $P_i$ can be evaluated using exactly one reordering with SS.

The sort key for FS is derived as follows. We choose a covering function of $C_{i,1}$ to be $\mathtt{wf}_{i,1}^*$. Let $\theta(P_i)$ denote the longest intersection among $\bigwedge_{\mathtt{wf}_j \in P_i}(\overrightarrow{\mathtt{WPK}_j} \circ \mathtt{WOK}_j)$ for every permutation $\overrightarrow{\mathtt{WPK}_j}$ of each $\mathtt{WPK}_j$. Since $P_i$ is prefixable, $\theta(P_i)$ is a sequence consisting of at least one attribute. Note that $\theta(P_i)$ might not be unique. As an example, in Example 8, $\theta(W)$ could be $abc$ or $bac$. By definition of $\theta(P_i)$, for each $\mathtt{wf}_j \in P_i$, there exists a permutation $\overrightarrow{\mathtt{WPK}_j}$ of $\mathtt{WPK}_j$ such that $\theta(P_i) \leq \overrightarrow{\mathtt{WPK}_j} \circ \mathtt{WOK}_j$.

Let $\gamma$ denote a covering permutation of $\mathtt{wf}_{i,1}^*$ such that $\theta(P_i) \leq \gamma$. Since $P_i$ is prefixable and $\mathtt{wf}_{i,1}^*$ is a covering function of $C_{i,1}$, $\gamma$ must exist. It follows that if we use $\gamma$ as the sort key for FS to reorder $\mathcal{I}_{i,1}$, Property 1 is guaranteed by the fact that $\gamma$ is a covering permutation of $\mathtt{wf}_{i,1}^*$, and Property 2 is guaranteed by the fact that $\theta(P_i) \leq \gamma$[5].

---
[5]Note that requiring $\theta(P_i) \leq \gamma$ is actually sufficient but not necessary for Property 2. Specifically, so long as $\theta' \leq \gamma$, where $\theta'$ is

### 4.5.2 Reordering with HS

We next discuss how to reorder $\mathcal{I}_{i,1}$ (wrt $\mathtt{wf}^*_{i,1}$) to $\mathcal{I}'_{i,1}$ using HS. Recall that HS is applicable if $\mathtt{WPK}_{i,1} \neq \emptyset$. Similar to FS, for HS, we need to choose the hash key, WHK, and sort key such that Properties 1 and 2 are satisfied.

Let $\theta'$ be the maximum prefix of $\theta(P_i)$ such that $attr(\theta') \subseteq \mathtt{WPK}_j$ for each $\mathtt{wf}_j \in C_{i,1}$. To satisfy Properties 1 and 2, it suffices to choose any subset of $\theta'$ for WHK. The selection of the sort key follows the same approach discussed in Section 4.5.1 for FS.

## 4.6 Further Optimization

In this section, we discuss some evaluation order issues related to our framework and discuss how our approach can be further optimized to mitigate the relation size assumption (Section 4.2).

Based on the preceding discussion, the evaluation plans produced by our evaluation framework (see Figure 2) are actually only partially ordered in that the evaluation order of some of the cover sets as well as window functions within a cover set could be reshuffled (without affecting correctness) for further optimization. Specifically, the following cover sets (csets) or window functions (wfs) can be reshuffled for further optimization: (1) the wfs within $C_0$, (2) the csets of $C_1$, (3) the $P_i$'s of $C_2$, (4) the csets of each $P_i$ in $C_2$, (5) the choice of covering functions for the first wf within each cset (except $C_0$), and (6) the non-first wfs within each $C_{i,j}$, $i \in [1,k], j \in [1,m_i]$.

To address the relation size assumption, one reasonable heuristic to reshuffle the above wfs/csets/$P_i$'s (refer to as units) is to order the units in increasing size of the extra column(s) produced by their evaluations so that the negative effect of a unit that is producing larger additional columns is deferred to a later stage of the execution chain. We intend to explore these further optimizations as part of our future work.

## 5. INTEGRATED WINDOW QUERY OPTIMIZATION

In this section, we present two approaches to integrate the optimization framework presented in the previous section into the overall query optimization process.

A window query *WQ* is essentially a conventional SQL statement $Q$ augmented with a set of window functions $W$ defined in the SELECT clause of $Q$. A *loose integration approach* to optimize *WQ* is to decompose the optimization task into a sequence of three optimization sub-tasks. First, optimize $Q$ (except for the DISTINCT and ORDER BY clauses) to produce a windowed table *WT*. Second, optimize the evaluation of $W$ on *WT* using the optimization framework in the previous section to produce an output table *WT'*. Finally, optimize the evaluation of the remaining DISTINCT and ORDER BY clauses on *WT'*.

While the loose integration approach offers a straightforward way to incorporate the window function optimization framework into a query optimizer, optimizing *WQ* as three separate sub-tasks can produce final query plans that are sub-optimal. For example, it is possible for a sub-optimal plan for the second sub-task to produce *WT'* ordered in an "interesting" order, which could lead to a less costly plan for the final sub-task and thus lead to an overall cheaper query plan.

This drawback can be alleviated by adopting a more *tightly integrated approach* to optimize *WQ*. Based on $W$ and $Q$, we identify a non-empty prefix of $\theta(P_i)$, Property 2 is guaranteed. However, the stronger requirement that we presented is beneficial for performance reason as the subsequent SS reorderings could be more efficient from sorting smaller segments.

a set *IP* of interesting order [16] and/or interesting grouping [15, 18] properties. Intuitively, *IP* consists of potential properties of *WT* that could benefit the derivation of *WT'* from *WT*. For example, suppose $Q$ contains a GROUP BY clause with a set of grouping attributes gpk. Then an interesting (order or grouping) property for *WT* would be for *WT* to be a segmented relation $WT^g_{\mathtt{gpk},\varepsilon}$ or $WT_{\emptyset, \overrightarrow{\mathtt{gpk}}}$ that would lead to a non-empty $C_0 \cup C_1$ for $W$. For each interesting property $ip$ in *IP*, the query optimizer will generate an optimal subplan to generate the windowed table $WT_{ip}$ that is associated with $ip$. In addition, the optimizer will also generate the optimal plan to produce an arbitrary windowed table $WT_o$ without taking into account of any interesting property in *IP*. Corresponding to each $WT_{ip}$ (or $WT_o$), we derive the optimal window function chain $C$ for evaluating $W$ on $WT_{ip}$ (or $WT_o$). Furthermore, by reshuffling the $P_i$'s of $C_2$ in $C$ (or the cover sets of $C_1$ if $C_2$ is empty), we also try to derive from $C$ the cheapest chain $C'$ that will result in a $WT'_{ip}$ (or $WT'_o$) (partially) satisfying the ordering requirement of the ORDER BY clause, so that an explicit sorting of $WT'_{ip}$ (or $WT'_o$) could be avoided or a cheaper partial sorting could be applicable. In this way, by taking into account of the interesting properties to enlarge the plan search space for *WQ*, the optimal query plan will not be missed.

## 6. PERFORMANCE STUDY

We validated our ideas using a prototype built in PostgreSQL 9.1.0 [1]. In our implementation, both the Hashed Sort (HS) and Segmented Sort (SS) are integrated into PostgreSQL as standard execution operators. Moreover, we modified the optimizer of PostgreSQL to support a total of four distinct optimization schemes, all of which generate window function chains as query plans:

CSO: Our proposed cover set-based optimization scheme.

BFO: The brute-force scheme, which enumerates and compares for a window query all the feasible execution plans that are based on FS, HS and SS.

ORCL: The scheme adopted by Oracle 8i [5]. It tries to cluster the window functions of a query into a minimum set of *Ordering Groups (OG)* which are equivalent to our notion of cover sets. However, the leading window function of each OG is only FS-reorderable.

PSQL: The naive scheme adopted by PostgreSQL 9.10, where the window functions of a query are evaluated strictly following their input sequence in the SELECT clause and each window function is only FS-reorderable. For a window function wf, the $\overrightarrow{\mathtt{WPK}}$ in the sort key ($\overrightarrow{\mathtt{WPK}} \circ \mathtt{WOK}$) of FS is exactly the input sequence of WPK attributes in the SELECT clause. The only optimization applied by PSQL is that, when a window function is matched by its input, the FS for it is omitted.

All experiments were performed on a Dell workstation with a 64-bit Intel Xeon X5355 2.66GHz processor, 4GB memory, one 500GB SATA disk and another 1TB SATA disk, running Linux 2.6.22. Both the operating system and PostgreSQL system are built on the 500GB disk, while the databases are stored on the 1TB disk.

### 6.1 Micro-benchmark Test on FS, HS, and SS

In this experiment, we used a micro-benchmark test to compare the performance of FS, HS and SS, under various situations. To this end, we defined a window query template $Q$:

```
SELECT *, rank() OVER
        (PARTITION BY AttrSet ORDER BY AttrSeq)
FROM T
```

$Q$ essentially evaluates a window rank() function with WPK = $AttrSet$ and WOK = $AttrSeq$ over a windowed table $T$, where

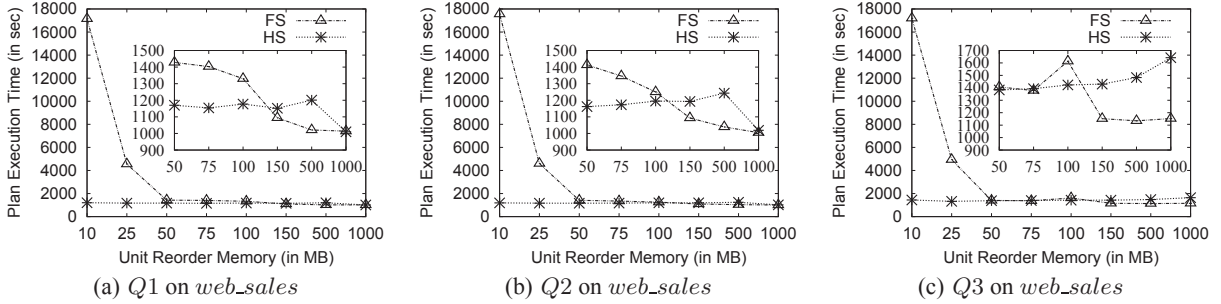

Figure 3: Micro-benchmark test, part 1: FS vs. HS

$AttrSet$, $AttrSeq$ and $T$ are all configurable. Another experimental parameter that we varied is the available operating memory dedicated to each tuple reordering operation, referred to as the *unit reorder memory* and denoted by $M$, whose values ranged from 10MB to 1000MB.

The execution of $Q$ invokes a single tuple reordering operation for `rank()`. For comparison, we directly measure the plan execution cost of $Q$, which consists of the cost of tuple reordering as well as the cost of the subsequent window function calls which usually remains constant.

| Part | Query | $T$ | $AttrSet$ | $AttrSeq$ |
|---|---|---|---|---|
| 1 | $Q1$ | $ws\_sales$ | $\{ws\_item\_sk\}$ | $(ws\_sold\_time\_sk)$ |
|   | $Q2$ |   | $\{ws\_item\_sk,$ $ws\_bill\_customer\_sk\}$ |   |
|   | $Q3$ |   | $\{ws\_warehouse\_sk\}$ |   |
| 2 | $Q4$ | $ws\_sales\_s$ | $\{ws\_quantity\}$ | $(ws\_item\_sk)$ |
|   | $Q5$ | $ws\_sales\_g$ |   |   |

Table 1: Queries used in the micro-benchmark test

We split the test into two parts. In the first part, $T$ was the $web\_sales$ relation from the TPC-DS [2] benchmark. We generated $web\_sales$ by using the official generator provided by TPC-DS and a scale factor of 100. The generated table had a total size of 14.3GB and contained 72 million tuples, each of which had the average tuple size of 214 bytes. All attributes in $web\_sales$ followed a uniform distribution. Since $web\_sales$ was totally unordered, the SS operation was inapplicable to reorder it. Thus, in this part we only compared the performance of FS and HS. We instantiated $Q$ with three concrete queries, $Q1$, $Q2$ and $Q3$ (shown in Table 1), which represent three kinds of situations where the number of window partitions are medium (204000), extremely large (71976736) and extremely small (16) respectively.

The experimental results are depicted in Fig. 3, from which we have several observations. First, while the performance of FS was sensitive to $M$ smaller than 150MB, relatively the performance of HS was very stable regardless of $M$. This is because the number of run merge passes in FS decreased from 6 to 1 as $M$ increased from 10MB to 150MB, while the sortings of hashed buckets in HS were either internal or external with a single run merge pass. Second, HS had huge (resp. decent) performance gains over FS when $M$ was smaller than 50MB (resp. between 50MB and 100MB), and lost out to FS when $M$ was larger than $100MB$. The reason for the performance loss of HS is that when $M \leq 150MB$, FS incurred just one pass of table I/O, but HS always incurred more than one pass of table I/O due to the table partitioning phase. However, we notice that in many situations, the performance loss of HS was negligible or insignificant. Third, we took a closer look into $Q3$, where each of the 16 window partitions had a large size of 900MB. Since we did not implement the optimization for HS, for each hashed bucket of HS in $Q3$, it always contained more than one window partition and thus had to be spilled to disk even when $M$ reached 1000MB. Thus, when $M$ increased, the I/O performance of HS did not improve, while the total CPU cost for tuple comparisons became higher and higher. This explains the performance reduction of HS along with the increased $M$ as shown in Fig. 3(c).

As a summary, HS is expected to outperform FS when $M$ is not very large. Moreover, another advantage of HS over FS is its performance stability under a very wide range of $M$s. On the other hand, a potential drawback of HS is that, unlike FS, its output does not have a total ordering, which may benefit the next stage's operations, e.g. an order by.

In the second part of this test, we compared the performance of SS with FS and HS. We generated two different instances of $T$: $web\_sales\_s$ and $web\_sales\_g$, both of which were manually reordered from the $web\_sales$ table used in the first part of this test. $web\_sales\_s$ (resp. $web\_sales\_g$) are sorted (resp. grouped) on attribute $ws\_quantity$. As such, we instantiated $Q$ with two concrete queries, $Q4$ and $Q5$ (shown in Table 1), for which the SS operation is applicable. Note that in both $Q4$ and $Q5$, SS will separately sort each $ws\_quantity$-group on $ws\_item\_sk$.

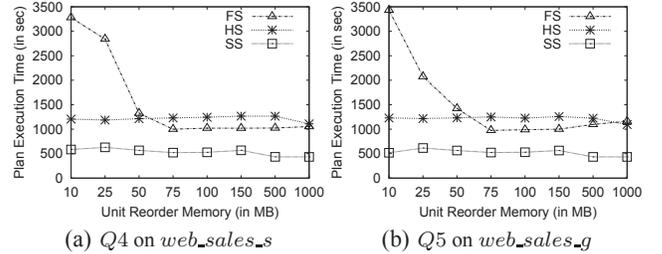

(a) $Q4$ on $web\_sales\_s$    (b) $Q5$ on $web\_sales\_g$

Figure 4: Micro-benchmark test, part 2: SS vs. FS and HS

As shown in Fig. 4, SS outperformed both FS and HS by a large margin in all situations. These experimental results are consistent with the prediction of the cost models.

### 6.2 Evaluation of Window Queries

In this experiment, we measured the effectiveness of our cover set-based window function optimization scheme presented in Section 4. To this end, we generated a list of window queries, each of which has the form of

SELECT *, $W$ FROM $web\_sales$

where $W$ represented a set of window `rank()` functions and $web\_sales$ (or $ws$ for short) was the TPC-DS table utilized in the first part of the above micro-benchmark test. The set of attributes of $ws\_sales$ mentioned by the tested window queries are listed in Table 2. For convenience, in the rest of this section we will refer to these attributes using their corresponding abbreviations.

The tested window queries were $Q6$, $Q7$, $Q8$ and $Q9$, whose embedded window functions are presented in Tables 3, 5, 7 and 9, respectively. Note that within each query, for two window functions $\mathtt{wf}_i$ and $\mathtt{wf}_j$, if $i < j$, then $\mathtt{wf}_i$ preceded $\mathtt{wf}_j$ in the SELECT clause. The optimized execution plans, i.e., window function chains, for the four tested window queries are shown in Tables 4, 6, 8 and 10, respectively. In a chain, $\mathtt{wf}_i/ws \to \mathtt{wf}_j$ (resp. $\mathtt{wf}_i/ws \xrightarrow{X} \mathtt{wf}_j$) means that the table $web\_sales$ or the output of $\mathtt{wf}_i$ matches (resp. needs to be reordered by $X = \mathtt{FS/HS/SS}$ for) $\mathtt{wf}_j$.

| Attribute | (abbr.) | Attribute | (abbr.) |
|---|---|---|---|
| $ws\_sold\_date\_sk$ | ($date$) | $ws\_item\_sk$ | ($item$) |
| $ws\_sold\_time\_sk$ | ($time$) | $ws\_bill\_customer\_sk$ | ($bill$) |
| $ws\_ship\_date\_sk$ | ($ship$) | | |

**Table 2: Attributes of $ws\_sales$ involved in the tested window queries, as well as their abbreviations**

We chose three values, 50MB, 75MB and 150MB, for the unit reorder memory $M$ allocated for each tuple reordering operation in a query plan. There are two reasons for 150MB being the maximum testing memory size. First, according to Fig. 3, in this experiment neither HS nor FS will strictly outperform the other. In so doing, we intended to examine the accuracy of our cost models for HS and FS proposed in Section 3.4. Second, it is observable from Fig. 3 and Fig. 4 that, compared with 150MB unit reorder memory, a larger memory size will make little difference on the performance of HS, FS and SS and thus will not invalidate the conclusions reached below.

We then compare the performance of CSO, WF, ORCL and PSQL according to each tested window query.

| | WPK | WOK | | WPK | WOK |
|---|---|---|---|---|---|
| $\mathtt{wf}_1$ | {$item$} | ($date$) | $\mathtt{wf}_2$ | {$item$} | ($bill$) |

**Table 3: Window functions contained by $Q6$**

| Scheme | $M$ (in MB) | Plan |
|---|---|---|
| BFO/CSO | 50/75 | $ws \xrightarrow{HS} \mathtt{wf}_1 \xrightarrow{SS} \mathtt{wf}_2$ |
| | 150 | $ws \xrightarrow{FS} \mathtt{wf}_1 \xrightarrow{SS} \mathtt{wf}_2$ |
| CSO(v1) | 50/75/150 | |
| CSO(v2) | 50/75 | $ws \xrightarrow{HS} \mathtt{wf}_1 \xrightarrow{HS} \mathtt{wf}_2$ |
| | 150 | $ws \xrightarrow{FS} \mathtt{wf}_1 \xrightarrow{FS} \mathtt{wf}_2$ |
| ORCL/PSQL | 50/75/150 | |

**Table 4: Execution plans for $Q6$**

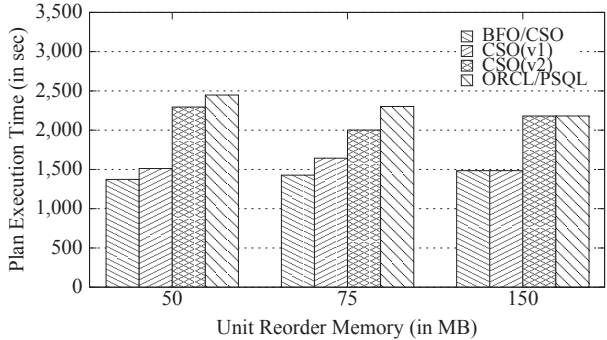

**Figure 5: Evaluating different optimization schemes with $Q6$**

For $Q6$, the execution plans generated by CSO (resp. PSQL) are exactly the same as those by BFO (resp. ORCL), as shown in Table 4. As such, we additionally tested two variants of CSO, i.e., CSO(v1) where HS is disabled, and CSO(v2) where SS is disabled. The experimental results are depicted in Fig. 5. It is obvious that, by assigning SS to $\mathtt{wf}_2$, BFO/CSO significantly improved the query performance of $Q6$. Moreover, when $M$ was 50MB/75MB, the introduction of CSO(v1) and CSO(v2) illustrated the cost difference between FS and HS for both $\mathtt{wf}_1$ and $\mathtt{wf}_2$. Thus, we can see that BFO/CSO made correct decisions on the choice between FS and HS for a window function, by using our proposed cost models.

| | WPK | WOK | | WPK | WOK |
|---|---|---|---|---|---|
| $\mathtt{wf}_1$ | {$date, time, ship$} | $\varepsilon$ | $\mathtt{wf}_4$ | $\emptyset$ | ($item, bill$) |
| $\mathtt{wf}_2$ | {$time, date$} | $\varepsilon$ | $\mathtt{wf}_5$ | {$date, time, item, bill$} | ($ship$) |
| $\mathtt{wf}_3$ | {$item$} | $\varepsilon$ | | | |

**Table 5: Window functions contained by $Q7$**

| Scheme | $M$ (in MB) | Plan |
|---|---|---|
| BFO | 50/75 | $ws \xrightarrow{HS} \mathtt{wf}_1 \to \mathtt{wf}_2 \xrightarrow{FS} \mathtt{wf}_5 \to \mathtt{wf}_4 \to \mathtt{wf}_3$ |
| | 150 | $ws \xrightarrow{FS} \mathtt{wf}_1 \to \mathtt{wf}_2 \xrightarrow{FS} \mathtt{wf}_5 \to \mathtt{wf}_4 \to \mathtt{wf}_3$ |
| CSO | 50/75 | $ws \xrightarrow{FS} \mathtt{wf}_5 \to \mathtt{wf}_4 \to \mathtt{wf}_3 \xrightarrow{HS} \mathtt{wf}_1 \to \mathtt{wf}_2$ |
| | 150 | $ws \xrightarrow{FS} \mathtt{wf}_5 \to \mathtt{wf}_4 \to \mathtt{wf}_3 \xrightarrow{FS} \mathtt{wf}_1 \to \mathtt{wf}_2$ |
| ORCL | 50/75/150 | |
| PSQL | 50/75/150 | $ws \xrightarrow{FS} \mathtt{wf}_1 \xrightarrow{FS} \mathtt{wf}_2 \xrightarrow{FS} \mathtt{wf}_3 \xrightarrow{FS} \mathtt{wf}_4 \xrightarrow{FS} \mathtt{wf}_5$ |

**Table 6: Execution plans for $Q7$**

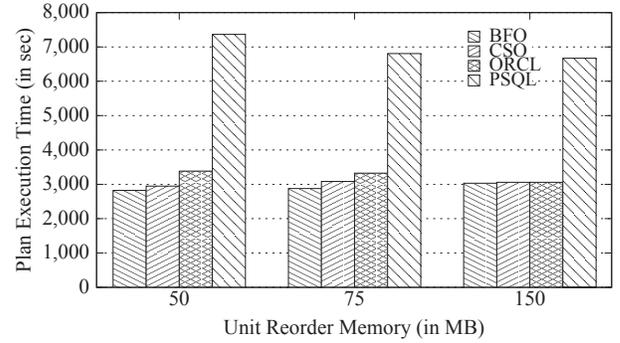

**Figure 6: Evaluating different optimization schemes with $Q7$**

$Q7$ is actually the running example utilized by [5] in order to illustrate the optimization mechanism of ORCL. As shown in Table 6, for this query BFO, CSO and ORCL all managed to find the execution plans with the same minimum set of cover sets. The differences between their plans lie in the choices between HS and FS for a single window function, as well as the sequence of window functions (or cover sets) in the chain. In contrast, $Q7$ actually highlighted the naiveness of PSQL, which failed to recognize the rather obvious optimization opportunity where adjusting the sort key of FS for $\mathtt{wf}_1$ is able to make the output of $\mathtt{wf}_1$ matches $\mathtt{wf}_2$. As a result, the performance of PSQL was much worse than that of BFO/CSO/ORCL, as depicted in Fig. 6. On the other hand, the small performance differences among BFO, CSO and ORCL gave rise to some observations. First, for $Q7$, once again both BFO and CSO made correct decisions on the choice between FS and HS for a window function, by using our proposed cost models. Second, when SS was not used, the sequence of cover sets in the chain has minor impact on the plan performance, and this is consistent with the intuition behind our *relation size assumption* made in Section 4.2.

$Q8$ was derived from $Q7$ by moving the $item$ attribute from $\mathtt{WOK}_4$ of $\mathtt{wf}_4$ into $\mathtt{WPK}_4$ and also moving the $bill$ attribute from $\mathtt{WPK}_5$ of $\mathtt{wf}_5$ into $\mathtt{WOK}_5$, according to Table 6 and Table 8. The resultant execution plans are listed in Table 8. We can see that each of BFO, CSO and ORCL generated a minimum but distinct set of cover sets. However, unlike BFO and CSO, ORCL cannot recognize the additional optimization opportunity that one of the three leading window functions of cover sets, i.e., $\mathtt{wf}_5$ here for ORCL, is actually SS-reorderable. Therefore, as shown in Fig. 7, the performance of ORCL was a bit worse than BFO and CSO, although PSQL still performed the worst. On the other hand, the performance difference between BFO and CSO was negligible, which showed that it is the numbers of cover sets and the SS, rather than their sequence in the chain, that affect the plan performance most significantly.

| | WPK | WOK | | WPK | WOK |
|---|---|---|---|---|---|
| $\text{wf}_1$ | $\{date, time, ship\}$ | $\varepsilon$ | $\text{wf}_4$ | $\{item\}$ | $(bill)$ |
| $\text{wf}_2$ | $\{time, date\}$ | $\varepsilon$ | $\text{wf}_5$ | $\{date, time, item\}$ | $(bill, ship)$ |
| $\text{wf}_3$ | $\{item\}$ | $\varepsilon$ | | | |

Table 7: Window functions contained by $Q8$

| Scheme | $M$ (in MB) | Plan |
|---|---|---|
| BFO | 50/75 | $ws \xrightarrow{\text{HS}} \text{wf}_1 \to \text{wf}_2 \xrightarrow{\text{SS}} \text{wf}_5 \xrightarrow{\text{HS}} \text{wf}_4 \to \text{wf}_3$ |
| | 150 | $ws \xrightarrow{\text{FS}} \text{wf}_1 \to \text{wf}_2 \xrightarrow{\text{SS}} \text{wf}_5 \xrightarrow{\text{FS}} \text{wf}_4 \to \text{wf}_3$ |
| CSO | 50/75 | $ws \xrightarrow{\text{HS}} \text{wf}_5 \xrightarrow{\text{SS}} \text{wf}_1 \to \text{wf}_2 \xrightarrow{\text{HS}} \text{wf}_4 \to \text{wf}_3$ |
| | 150 | $ws \xrightarrow{\text{FS}} \text{wf}_5 \xrightarrow{\text{SS}} \text{wf}_1 \to \text{wf}_2 \xrightarrow{\text{FS}} \text{wf}_4 \to \text{wf}_3$ |
| ORCL | 50/75/150 | $ws \xrightarrow{\text{FS}} \text{wf}_4 \to \text{wf}_3 \xrightarrow{\text{FS}} \text{wf}_5 \to \text{wf}_2 \xrightarrow{\text{FS}} \text{wf}_1$ |
| PSQL | 50/75/150 | $ws \xrightarrow{\text{FS}} \text{wf}_1 \xrightarrow{\text{FS}} \text{wf}_2 \xrightarrow{\text{FS}} \text{wf}_3 \xrightarrow{\text{FS}} \text{wf}_4 \xrightarrow{\text{FS}} \text{wf}_5$ |

Table 8: Execution plans for $Q8$

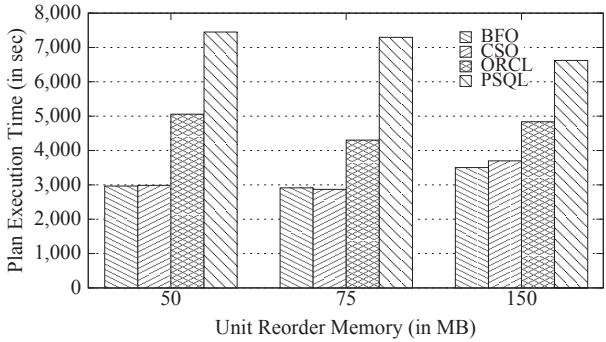

Figure 7: Evaluating different optimization schemes with $Q8$

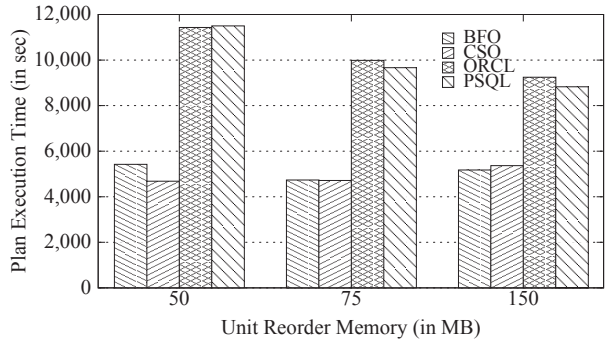

Figure 8: Evaluating different optimization schemes with $Q9$

The final tested query $Q9$ contained the most window functions and thus led to the most complicated execution plans, which are presented in Table 10. This time PSQL eventually succeeded in avoiding the FS for $\text{wf}_3$ and thus generated an execution plan that was comparable with that of ORCL. The reasons that ORCL lost out to BFO and CSO to a very large extent are two-fold. On the one hand, ORCL generated one more cover set than BFO and CSO; on the other hand, in essence it was not able to recognize those optimization opportunities w.r.t SS. As for BFO and CSO, Table 10 and Fig. 8 together show that their optimization effects were equally well. The fact that CSO outperformed BFO with 50MB memory was due to the accidental inconsistency between the actual plan execution costs and our cost model's estimations.

| | WPK | WOK | | WPK | WOK |
|---|---|---|---|---|---|
| $\text{wf}_1$ | $\{item\}$ | $(bill, date)$ | $\text{wf}_5$ | $\{bill, date\}$ | $(time)$ |
| $\text{wf}_2$ | $\{item, time\}$ | $(date)$ | $\text{wf}_6$ | $\{bill\}$ | $(time)$ |
| $\text{wf}_3$ | $\{item\}$ | $(time)$ | $\text{wf}_7$ | $\{date, time\}$ | $\varepsilon$ |
| $\text{wf}_4$ | $\emptyset$ | $(item, date)$ | $\text{wf}_8$ | $\emptyset$ | $(time)$ |

Table 9: Window functions contained by $Q9$

| Scheme | $M$ (in MB) | Plan |
|---|---|---|
| BFO | 50/75 | $ws \xrightarrow{\text{FS}} \text{wf}_1 \xrightarrow{\text{FS}} \text{wf}_2 \to \text{wf}_3 \xrightarrow{\text{SS}} \text{wf}_4$ $\xrightarrow{\text{HS}} \text{wf}_5 \xrightarrow{\text{SS}} \text{wf}_6 \xrightarrow{\text{FS}} \text{wf}_7 \to \text{wf}_8$ |
| | 150 | $ws \xrightarrow{\text{FS}} \text{wf}_1 \xrightarrow{\text{SS}} \text{wf}_2 \to \text{wf}_3 \xrightarrow{\text{SS}} \text{wf}_4$ $\xrightarrow{\text{FS}} \text{wf}_5 \xrightarrow{\text{SS}} \text{wf}_6 \xrightarrow{\text{FS}} \text{wf}_7 \to \text{wf}_8$ |
| CSO | 50/75 | $ws \xrightarrow{\text{FS}} \text{wf}_7 \to \text{wf}_8 \xrightarrow{\text{HS}} \text{wf}_6 \xrightarrow{\text{SS}} \text{wf}_5$ $\xrightarrow{\text{FS}} \text{wf}_2 \to \text{wf}_3 \xrightarrow{\text{SS}} \text{wf}_1 \xrightarrow{\text{SS}} \text{wf}_4$ |
| | 150 | $ws \xrightarrow{\text{FS}} \text{wf}_7 \to \text{wf}_8 \xrightarrow{\text{FS}} \text{wf}_6 \xrightarrow{\text{SS}} \text{wf}_5$ $\xrightarrow{\text{FS}} \text{wf}_2 \to \text{wf}_3 \xrightarrow{\text{SS}} \text{wf}_1 \xrightarrow{\text{SS}} \text{wf}_4$ |
| ORCL | 50/75/150 | $ws \xrightarrow{\text{FS}} \text{wf}_2 \to \text{wf}_8 \xrightarrow{\text{FS}} \text{wf}_4 \xrightarrow{\text{FS}} \text{wf}_7$ $\xrightarrow{\text{FS}} \text{wf}_1 \xrightarrow{\text{FS}} \text{wf}_3 \xrightarrow{\text{FS}} \text{wf}_6 \xrightarrow{\text{FS}} \text{wf}_5$ |
| PSQL | 50/75/150 | $ws \xrightarrow{\text{FS}} \text{wf}_1 \xrightarrow{\text{FS}} \text{wf}_2 \to \text{wf}_3 \xrightarrow{\text{FS}} \text{wf}_4$ $\xrightarrow{\text{FS}} \text{wf}_5 \xrightarrow{\text{FS}} \text{wf}_6 \xrightarrow{\text{FS}} \text{wf}_7 \xrightarrow{\text{FS}} \text{wf}_8$ |

Table 10: Execution plans for $Q9$

In a summary, BFO and CSO always delivered the best execution plans for all the four tested queries, $Q6$, $Q7$, $Q8$ and $Q9$. ORCL performed much worse than BFO and CSO but significantly better than PSQL at the same time.

## 6.3 Optimization Overheads

In this experiment, we subsequently compared the optimization overheads of those four optimization schemes. We generated a list of window queries with different number of window functions to be evaluated on the $web\_sales$ table. In each window function wf of each query, we randomly determined the number of attributes as well as the attributes themselves for both WPK and WOK. The overheads of different optimization schemes incurred by optimizing six queries, where the number of window functions ranged from 6 to 10, are listed in Table 11.

| Scheme \ # of wfs | 6 | 7 | 8 | 9 | 10 |
|---|---|---|---|---|---|
| BFO | 1.56 | 18.47 | 9336 | 489286 | $9.8 \times 10^6$ |
| CSO | 1.44 | 3.56 | 4.07 | 7.49 | 12.31 |
| ORCL | 0.99 | 1.04 | 1.27 | 1.36 | 1.49 |
| PSQL | 0.85 | 0.91 | 1.03 | 1.11 | 1.18 |

Table 11: Optimization overheads (in millisecond) of optimization schemes for queries with varying numbers of window functions

From Table 11 we can see that, the optimization overheads of both ORCL and PSQL increased slowly along with the number of window functions. As for CSO, its optimization overheads increased a bit faster but still remained very small. However, for BFO, as expected, its optimization overheads were acceptable for upto 7 window functions but became totally unacceptable when the number of window functions exceed 8. In particular, BFO took about 2.7 hours to derive the optimal plan for a query with 10 window functions!

According to the experimental results of the last experiment, the effectiveness of CSO is very similar to that of BFO which is supposed to always generate optimal plans. Moreover, as illustrated above, CSO is much lighter-weight than BFO. As such, we can conclude that CSO achieves the best tradeoff between optimization effectiveness and optimization efficiency.

## 7. RELATED WORK

To the best of our knowledge, [5] is the only research report in the public domain that focuses on optimizing the evaluation of window functions. That work only used `FS` for tuple reordering. In contrast, we propose two new tuple reordering operations `HS` and `SS`, both of which are competitive alternatives to `FS`. The optimization scheme proposed in [5] also exploited the properties of `WPK`s and `WOK`s, and clusters window functions into *Ordering Groups* which are equivalent to our notion of *cover sets*, so as to minimize the total number of `FS` operations needed. However, our cover set-based optimization scheme naturally subsumes the optimization scheme of [5] and incorporates additional optimizations w.r.t `HS` and `SS`. In addition, the other two window function optimizations mentioned in [5], i.e. predicate pushdown for ranking functions and parallel execution of a single window function, are both complementary and can co-exist with our approaches.

The window functions in a window query compute a set of additional window-function columns for a single windowed table, according to different specifications of window partitioning and ordering. Similarly, `GROUPING SETS`, `ROLLUP` and `CUBE` operations, the three extensions to the `GROUP BY` clause, group the tuples of a table in multiple disparate ways, compute aggregations for different tuple groups, and finally concatenate all of the tuple groups together into a single result[6]. However, the optimization techniques available for these `GROUP BY` extensions (e.g. [3], [10] and [11]) cannot be directly applied for window function evaluation. First, the window function evaluation retains the original table tuples, while in the evaluation of these `GROUP BY` extensions, each tuple group collapses into a single tuple. Second, the window functions contain in a window query can be of different types, while in these `GROUP BY` extensions, a set of global aggregation functions are evaluated for different tuple groups.

Research works like [15, 17, 18] have proposed optimization frameworks to infer the ordering and grouping properties held for intermediate results of query execution by using functional dependencies. They aimed at avoiding redundant sort and group operations in the query plan. In this paper, we also need to infer the properties of the intermediate results flowing between window functions, in order to determine the proper tuple reordering operations. However, the only interesting property that we formulate is the relation segmentation as described in Section 3, of which both ordering and grouping are special cases. Moreover, in our framework, the operations that can alter the tuple ordering include the newly proposed `HS` and `SS`, whose behaviors are significantly different from those operations considered in [15, 18], such as group by and join. As a result, their techniques are not directly applicable or extendible in our problem context.

## 8. CONCLUSION

In this paper, we have presented a comprehensive framework for optimizing the evaluation of window functions. We have proposed two new tuple reordering methods, namely Hashed Sort (`HS`) and Segmented Sort (`SS`), that can efficiently reorder tuples for window function evaluation. To handle complex queries involving multiple window functions, we also designed a light-weight cover set-based optimization scheme that generates a (near-)optimal window function chain for evaluating these window functions. We have integrated our techniques into PostgreSQL. Our extensive performance study showed that our techniques can bring substantial performance gains over existing window function implementations and optimizations.

There are several directions for future work. First, the functional dependencies existing between attributes of the windowed table have non-trivial impact on the optimizations for window functions, and thus should be exploited further. Second, it is possible to develop some tailored optimizations for certain types of window functions, like the predicate pushdown optimization for window ranking functions as proposed in [5]. We plan to investigate other kinds of window functions to identify additional optimizations for them. Finally, in this paper we assume a sequential evaluation model for window functions. However, an alternative evaluation model is a graph-based evaluation model, where a window function may receive input from multiple sources and may deliver its output to multiple destinations. Studying the effectiveness of such a graph-based model is certainly in our agenda for future work.

---

[6]Note that both `ROLLUP` and `CUBE` are special cases of `GROUPING SETS`.